\documentclass[conference]{IEEEtran}
\IEEEoverridecommandlockouts

\usepackage{cite}
\usepackage{amsmath,amssymb,amsfonts}
\usepackage{graphicx}
\usepackage{epstopdf}
\usepackage{subfigure}
\usepackage{stfloats}
\usepackage{textcomp}
\usepackage{xcolor}
\usepackage{flushend}
\usepackage{array}
\usepackage{algorithm}
\usepackage{algpseudocode}
\usepackage{multirow}
\usepackage{amsthm}

\newtheorem{proposition}{Proposition}

\def\BibTeX{{\rm B\kern-.05em{\sc i\kern-.025em b}\kern-.08em
    T\kern-.1667em\lower.7ex\hbox{E}\kern-.125emX}}

\begin{document}

\title{Antenna Selection for Enhancing Privacy in Radar-Based Vital Sign Monitoring Systems
\thanks{This work was supported by ARO grant W911NF2320103 and NSF grant ECCS-2320568.}
}

\author{\IEEEauthorblockN{Zhihao Tao}
\IEEEauthorblockA{\textit{Dept. of Electrical and Computer Engineering} \\
\textit{Rutgers, the State University of New Jersey}\\
New Brunswick, USA \\
zt118@scarletmail.rutgers.edu}
\and
\IEEEauthorblockN{Athina P. Petropulu}
\IEEEauthorblockA{\textit{Dept. of Electrical and Computer Engineering} \\
\textit{Rutgers, the State University of New Jersey}\\
New Brunswick, USA \\
athinap@soe.rutgers.edu}
}

\maketitle

\begin{abstract}
Radar-based vital sign monitoring (VSM) systems have become valuable for non-contact health monitoring by detecting physiological activities, such as respiration and heartbeat, remotely. However, the conventional phased array used in VSM is vulnerable to privacy breaches, as an eavesdropper can extract sensitive vital sign information by analyzing the reflected radar signals. In this paper, we propose a novel approach to protect privacy in radar-based VSM by modifying the radar transmitter hardware, specifically by strategically selecting the transmit antennas from the available antennas in the transmit array.
By dynamically selecting which antennas connect or disconnect to the radio frequency chain, the transmitter introduces additional phase noise to the radar echoes, generating false frequencies in the power spectrum of the extracted phases at the eavesdropper's receiver. 
The antenna activation pattern is designed to maximize the variance of the phases introduced by antenna selection, which effectively makes the  false frequencies dominate the spectrum, obscuring the actual vital sign frequencies. 
Meanwhile, the authorized receiver, having knowledge of the antenna selection pattern, can compensate for the phase noise and accurately extract the vital signs. Numerical experiments are conducted to validate the effectiveness of the proposed approach in enhancing privacy while maintaining vital sign monitoring.
\end{abstract}

\begin{IEEEkeywords}
Antenna selection, phased array radar, privacy protection, vital sign monitoring.
\end{IEEEkeywords}

\section{Introduction}
Remote vital sign monitoring (VSM) has been a topic of growing interest since the 1970s due to its ability to continuously and non-intrusively monitor heart rate and respiratory rate, which are here referred to as  physiological parameters  \cite{lin1975noninvasive,wang2020review}. Remote VSM methods have leveraged a variety of signals, including WiFi and IoT communications, where fluctuations in the wireless channel provide insights into the monitored individual’s movements and vital signs \cite{baig2014real, lopez2023comprehensive, wang2024prisense}. Active radar-based VSM has also been extensively studied, focusing on extracting subtle body vibrations associated with vital sign activities from radar echoes \cite{gu2013hybrid,nosrati2019concurrent,li2013review,xu2022simultaneous}. Among these approaches, phased array radar is particularly promising due to its widespread applicability, and cost-effectiveness \cite{xu2023flexible,xu2024phased}. Phased array radar typically transmit either continuous-wave (CW) or linear frequency-modulated continuous-wave (FMCW) signals for VSM \cite{li2013review,xu2023flexible,xu2024phased,mercuri2017frequency,ahmad2018vital,xu2022simultaneous}. Compared to traditional VSM, which requires attaching sensors to the body, radar-based VSM is particularly suitable for applications where hygiene, continuous monitoring, and minimal discomfort are crucial, such as monitoring driver drowsiness, sleep quality, and infant cardiac health \cite{gharamohammadi2023vehicle, hussain2022non,singh2020multi}.

Previous research on radar-based VSM has primarily focused on advancing monitoring hardware, developing beamforming techniques and signal processing algorithms \cite{yuan2020high, xu2023flexible, xu2024phased, hall2015phased, rahman2015signal}, with limited attention given to the privacy of the vital sign data. However, vital signs contain sensitive health information and can even serve as unique identifiers for individuals \cite{tan2022commodity}. Ensuring user privacy in remote VSM applications is therefore crucial. In VSM methods that utilize WiFi signals \cite{baig2014real}, privacy concerns have been addressed by implementing network authentication techniques that restrict access to authorized devices only. Similarly, privacy in IoT-based VSM \cite{wang2024prisense} has been safeguarded by transmitting information that distorts the channel of the eavesdroppers, thus preventing them from extracting vital sign details. To the best of the authors' knowledge, privacy protection for radar-based VSM was first addressed in \cite{tao2024nonlinear}, where vital sign privacy is protected through a waveform design approach, specifically by using non-linear FMCW waveforms to modulate the phases of the radar signals. By properly designing the waveform parameters, the approach of \cite{tao2024nonlinear}  makes it more difficult for the eavesdropper to decipher vital signs while allowing the authorized receiver to extract them easily.
However, such nonlinear waveform-based method requires precise range and timing information, introducing extra communication overhead. Integrating it into existing phased array radar hardware can also be challenging due to the added complexity in synchronization and signal processing requirements.

In this paper, we take a hardware-oriented approach to address privacy challenges in radar-based VSM by focusing on  transmitter design. Specifically, we propose an antenna selection-based method that strategically activates and deactivates specific antennas in the phased array over time, thus introducing time-varying phases into the reflected radar signals. The time-varying phases endow the transmitter waveforms with additional
phase noises and can generate false frequencies over the power spectrum of eavesdropper's extracted phases. To effectively obscure the actual vital sign frequencies, the power of the false frequencies must exceed that of the true frequencies. This can be achieved by maximizing the variance of the introduced phases through antenna selection. We propose two antenna selection patterns to maximize the phase variance. One involves optimizing the number of activated antennas and the antenna weight vector simultaneously, while the other focuses on optimizing the probability of each phase occurrence. Moreover, the authorized radar receiver, with knowledge of the antenna selection pattern, can easily compensate for the additional phase noise introduced by the transmitter. This allows the authorized receiver to accurately reconstruct the vital signs without degradation of monitoring quality. Numerical experiments are conducted to validate the effectiveness of our proposed antenna selection-enabled transmitter, demonstrating its capability to enhance privacy while implementing vital sign monitoring. In addition, antenna selection can be realized by just using single-pole-single-throw (SPST) switches, which are low-cost and compatible well with the radar transceiver, so, by leveraging a hardware-based solution, our approach integrates seamlessly with existing phased array radar systems that transmit CW or linear FMCW waveforms.

Antenna selection was also used in \cite{valliappan2013antenna, rocca20134} to reduce the sidelobes in directions other than the intended receiver, and thus  improve physical layer security. 
The works of \cite{valliappan2013antenna, rocca20134} require knowledge of the direction of the eavesdropper, and in the absence of such knowledge, they scan all possible directions to optimize the antenna selection scheme. However, the latter approach increases the design complexity. Different than \cite{valliappan2013antenna, rocca20134}, our proposed approach customizes the antenna activation pattern to generate artificial frequencies at the transmitter, confounding the eavesdropper when it attempts to extract true vital sign information from the frequency spectrum. This design is independent of the receiver and does not require knowledge of the eavesdropper's direction.


The remainder of the paper is organized as follows. Section II describes how an eavesdropper could decipher vital signs in a conventional phased array radar-based VSM system. Section III illustrates how antenna selection works to enhance privacy while monitoring vital signs. Two antenna selection patterns are designed to maximize phase variance, ensuring that the generated false frequencies obscure the actual vital sign information. Section IV and Section V include numerical results and conclusions, respectively.

\section{Conventional Phased Array Radar-Based VSM Systems}\label{sec2}

\begin{figure}[t]
\centerline{\includegraphics[width=2.6in]{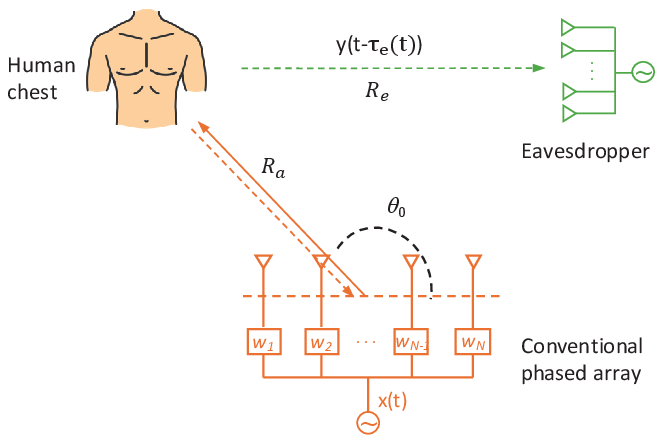}}
\caption{A conventional phased array radar-based VSM system with an eavesdropper.}\label{fig1}
\end{figure}

Consider a phased array radar-based VSM system in the presence of an eavesdropper, as shown in Fig. \ref{fig1}. The nominal distance between the authorized radar receiver and the user being monitored (to be referred to in the following as ``user")  is denoted by $R_a$, and the distance between the eavesdropping receiver and the user by $R_e$. The user is located at the direction $\theta_0$ with respect to the radar transmitter. The chest displacement,  $R(t)$, varying in time, $t$, due to  vital sign activities such as  breathing and heartbeat, can be approximated as \cite{xu2023flexible, xu2024phased, mercuri2017frequency}
\begin{equation}
R(t) \approx A_h \sin (2 \pi f_h t + \phi _h) + A_b \sin (2 \pi f_b t + \phi _b),
\end{equation}
where $\{A_h, f_h, \phi _h\}$ and $\{A_b, f_b, \phi _b\}$ are the corresponding amplitude, frequency, and initial phase of heartbeat and breathing, respectively. The range of $f_h$ is $0.8-2$ Hz and the range of $f_b$ is $0.1-0.5$ Hz  \cite{ti2024sensors}. Considering a CW transmit waveform, the transmitted baseband signal is
\begin{equation}
    x(t) = A_0 e^{j (2 \pi f_c t + \phi_0)},
\end{equation}
where $A_0$ and $\phi_0$ are the waveform amplitude and initial phase. For a phased array containing $N$ elements, the radiated radar signal into the half-space, $\theta \in [0, \pi]$, can be expressed as
\begin{equation}\label{eq1}
    y(t, \theta) = x(t) \sum _{n=1}^{N} w_n e^{j \frac{2\pi}{\lambda} (n-1) d \cos \theta},
\end{equation}
where $w_n$ is the antenna weight at the $n$-th array element, $\lambda$ is the wavelength and $d$ is the inter-element distance. $d$ is usually set as $\lambda /2$ and $w_n = e^{-j (n-1) \pi \cos \theta_c}$, $\theta_c \in [0, \pi]$. When $\theta_c = \theta_0$, the radiated beam is focused on the  object being monitored,  so that the  signal towards the user is $y(t) = Nx(t)$. In all other directions, $\theta_c \ne \theta_0$,  the radiated signal appears as 
\begin{equation}\label{eq2}
\begin{split}
    y(t;{\theta_0,\theta_c}) &= x(t) \sum _{n=1}^{N} e^{j (n-1) \pi (\cos \theta_0 - \cos \theta_c)}\\
    &= x(t) \frac{\sin(\frac{N}{2}\pi (\cos \theta_0 - \cos \theta_c))}{\sin(\frac{1}{2}\pi (\cos \theta_0 - \cos \theta_c))} e^{j \frac{N-1}{2} \pi (\cos \theta_0 - \cos \theta_c)}\\
    &= A_0A_p(\theta_0, \theta_c) e^{j (2 \pi f_c t + \phi_0 + \phi _p(\theta_0, \theta_c))},
\end{split}
\end{equation}
where $A_p = \frac{\sin(\frac{N}{2}\pi (\cos \theta_0 - \cos \theta_c))}{\sin(\frac{1}{2}\pi (\cos \theta_0 - \cos \theta_c))}$ and $\phi _p = \frac{N-1}{2} \pi (\cos \theta_0 - \cos \theta_c)$ are the fixed amplitude and phase introduced by the phased array. Since $A_p(.) \rightarrow N$ and $\phi_p \rightarrow 0$ when $\theta_c \rightarrow \theta _0$, we use \eqref{eq2} to represent the radiated signal towards the user for $\forall \theta_c \in [0, \pi]$.

The radiated signal is then reflected by the user, and is received by the eavesdropper as $r(t) = \alpha y(t - \tau _e (t);{\theta_0,\theta_c}) + n(t)$, where $\alpha$ accounts for the propagation loss, the target radar cross section and the reflection effects, $n(t)$ consists of receiver noises and
\begin{equation}
    \tau _e (t) = \frac{R_a + R_e + 2R(t)}{c},
\end{equation}
$c$ is the light speed. We assume that the eavesdropper knows the carrier frequency of the radar transceiver, $f_c$, which is not difficult to acquire. Here, we assume that intrinsic phase noise, receiver noise, propagation loss, and reflection have negligible effects. If these effects were significant, they would inherently protect the user's vital signs, eliminating the need for additional privacy measures.

With knowledge of  $f_c$, the eavesdropper can mix the received signal with  $e^{-j 2 \pi f_c t}$ to  obtain 
{\small {\begin{equation}\label{eq3}
\begin{split}
    s(t;{\theta_0,\theta_c}) &= \alpha A_0 A_p(\theta_0, \theta_c) e^{-j 2 \pi f_c (\frac{R_a + R_e}{c} + \frac{2R(t)}{c}) + j(\phi_0 + \phi _p(\theta_0, \theta_c))}\\
    &= \alpha A_0 A_1 A_p(\theta_0, \theta_c) e^{-j (\frac{4 \pi}{\lambda}R(t)-\phi _p(\theta_0, \theta_c))},
\end{split}
\end{equation}}} 
where $A_1 = e^{-j 2 \pi f_c \frac{R_a + R_e}{c} + j \phi_0}$. Here we assume that the receiver is sensitive enough so it can extract the continuous phases of the mixed signal. Since $\alpha, A_0, A_1, A_p(.)$ and $\phi_p$ are all constant amplitude or phase, the eavesdropper can take the following steps to decipher the vital signs:  extract the unwrapped phase   of the mixed signal,  remove its mean, and then   take a discrete Fourier transform (DFT) on the demeaned phase.The locations of the DFT peaks would reveal  the  vital sign frequencies in $R(t)$.

\section{Antenna Selection-Enabled Phased Array for Enhancing VSM Privacy}

\subsection{Antenna Selection-Enabled Transmitter}
\begin{figure}[t]
\centerline{\includegraphics[width=2.6in]{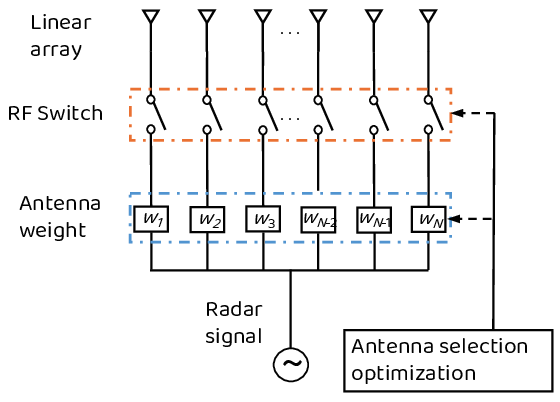}}
\caption{An antenna selection-enabled phased array transmitter.}\label{fig2}
\end{figure}

From \eqref{eq3}, it follows that the phase $\phi_p$ introduced by the conventional phased array remains constant over time in the user’s direction. Consequently, it does not obscure the phase information in the reflected signal, allowing an eavesdropper to easily intercept the target user's vital sign frequencies.
To mitigate this vulnerability, we propose an antenna selection scheme that dynamically activates a subset of antennas in the phased array at each time instance using a series of SPST switches, as shown in Fig. \ref{fig2}. Modern RF switches operate on the nanosecond to microsecond scale; here, we set the switching intervals to match the sampling rate of the receivers.
The radiated signal towards to the user from an antenna selection-enabled phased array transmitter can be expressed as
\begin{equation}
    y(t;{\theta_0,\theta_c}) = x(t) \sum_{n=1}^{N} b_n(t) w_n(\theta_c) e^{j \pi (n-1) d \cos \theta_0},
\end{equation}
where $b_n(t)$ is a binary variable indicating the activation status of the \(n\)-th antenna at time \(t\). $b_n(k) = 1$ if the antenna is active and otherwise it is 0. When denoting $N$ antenna activation statuses as a vector $\boldsymbol{b}(t) = [b_1(t),\cdots,b_n(t)]^T$, one can get
\begin{equation}
    y(t;{\theta_0,\theta_c}) = x(t)\boldsymbol{b}^T(t) \boldsymbol{f}(\theta_0,\theta_c),
\end{equation}
where $\boldsymbol{f}(\theta_0, \theta_c) = [1, e^{j \pi \cos (\theta_0 - \theta_c)}, \cdots, e^{j (N-1) \pi \cos (\theta_0 - \theta_c)}]^T$. The received signal at the eavesdropper is $r(t) = \alpha x(t - \tau _e (t)) \boldsymbol{b}(t - \tau _e (t)) \boldsymbol{f}(\theta_0,\theta_c)$. Since the time delay $\tau _e (t)$ is very small, $\boldsymbol{b}(t - \tau _e (t))$ can be approximated by $\boldsymbol{b}(t)$ when the antenna activation pattern change slowly. After mixing, the eavesdropper obtains
\begin{equation}\label{eq4}
    s(t;{\theta_0,\theta_c}) = \alpha A_0 A_1 A_p(t,\theta_0, \theta_c) e^{-j (\frac{4 \pi}{\lambda}R(t)-\phi _p(t,\theta_0, \theta_c))},
\end{equation}
where
\begin{equation}\label{eq_angle}
    \begin{split}
        A_p(t,\theta_0, \theta_c) &= \left |\boldsymbol{b}^T(t) \boldsymbol{f}(\theta_0, \theta_c)\right |,\\
        \phi _p(t,\theta_0, \theta_c) &= \arg(\boldsymbol{b}^T(t) \boldsymbol{f}(\theta_0, \theta_c)) \\&= \arg(\sum_{n=1}^N b_n(t)e^{j\pi (n-1) cos(\theta_0-\theta_c)}).
    \end{split}
\end{equation}

\subsection{Enhanced Privacy via Maximization of the Phase Variance}\label{sec3b}
We can see from \eqref{eq4} that the activated subset of antennas varies over time, introducing a time-varying phase shift $\phi _p(t,{\theta_0,\theta_c})$, which appears as randomized phase noise in the power spectrum. This time-varying phase, if designed properly, can  generate false vital sign frequencies to confound the eavesdropper when it uses spectral estimation methods to find the true vital sign frequencies. To this end, one might consider introducing sinusoidal variations in \( \phi_p(t,{\theta_0,\theta_c}) \) to produce predictable false frequencies. However, the discrete nature of antenna configurations, limited by \(M=\binom{N}{L} \) ($L$ is the number of activated antennas at each time), restricts the available phase values. These limited configurations make it impractical to achieve a smooth sinusoidal variation over time. Instead, maximizing the variance of \( \phi_p(t,{\theta_0,\theta_c}))\) spreads the introduced phase noise across a broad range of frequencies in a randomized manner, making it more challenging for an eavesdropper to isolate and interpret the physiological information.
We propose two schemes as follows to manipulate the phases.

\medskip
\noindent \textit{Optimization of Antenna Activation Parameters}: 
Let the sampling time be $T_s$ and the sampling frequency of the receiver be $f_s$, so the number of total sampling points is $S = T_s f_s$. At each sampling point, $t_j$, $L$  elements out of $N$ total elements are selected to be active, giving rise to   \(M=\binom{N}{L} \) possible antenna configurations; each configuration corresponds to a different \({\textbf b}(t)\) vector, and its  phase shift \(\phi_p(t_j,\theta_0,\theta_c)\) (see \eqref{eq_angle}).
%
Thus, at each sampling time point  
the introduced phase shift takes one of $M$ possible the values; let us denote those values as $\{\phi_i, i=1,...,M\}$. Then we adopt a uniform distributed antenna selection pattern, i.e.,  a uniform selection of the corresponding phases with equal probability  $1/M$.
%
Let us also determine the  number of activated antennas $L$  to maximize the variance of the introduced phase sequence, i.e.,
\begin{equation}\label{eq5}
\begin{split}
    \underset{L}{\text{max}}& \quad \text{Var}(\phi_p(t_j,\theta_0,\theta_c)_{j=1,\cdots,S}) \\
    \text{s.t.}& \quad 0 \leq \theta_c \leq \pi,\\
                &\quad 1 \leq L \leq N-1.
\end{split}
\end{equation}
Considering that \eqref{eq5} is a mixed-integer nonlinear programming problem, we can use heuristic methods to solve it.



\medskip
\noindent \textit{Optimization of Probability Allocation of Antenna Configurations}:  Here, we consider different probabilities, $p_i$, for each antenna configuration, or equivalently, of each phase, and maximize the phase variance by adjusting the probability distribution $p_i$ of each phase. For a fixed $\theta _c$ and $L$, we can formulate the following constraint optimization problem:
\begin{equation}
\begin{split}
    \underset{p_i}{\text{max}}& \quad \text{Var}(\phi_p(t_j,\theta_0,\theta_c)_{j=1,\cdots,S}) \\
    \text{s.t.}& \quad \sum_{i=1}^{M} p_i = 1 \\
                &\quad p_i \geq 0 \quad \forall i \in [1,M],
\end{split}
\end{equation}
where
\begin{equation}\label{eq6}
    \begin{split}
        \text{Var}(\phi_p(t),\theta_0,\theta_c))
        &= \sum_{i=1}^M p_i \phi_i^2 - \left( \sum_{i=1}^M p_i \phi_i \right)^2.
    \end{split}
\end{equation}
Solving this problem leads to the following proposition, which provides a closed-form solution for the optimal probability allocation in the context of maximizing the phase variance.
\begin{proposition}
    Let $\phi_{\text{max}}$ and $\phi_{\text{min}}$ be respectively the maximal and minimal phases among $\{\phi_i\}_{i=1,\cdots,M}$. The phase variance is maximized when only two configurations $\phi_a$ and $\phi_b$ are assigned non-zero probabilities $p_a$ and $1-p_a$, respectively, and when $\phi_a = \phi_{\text{min}}$, $\phi_b = \phi_{\text{max}}$, and $p_a = \frac{1}{2}$. The resulting maximum phase variance is:
    \begin{equation}
        \text{Var}(\phi_p(t))) = \frac{1}{4} (\phi_{\text{max}} - \phi_{\text{min}})^2.
    \end{equation}
\end{proposition}
The proof is shown in the Appendix A.

\subsection{Extract Vital Signs via Known Antenna Selection Patterns}
The received radar echo by the authorized receiver can be written as $r(t) = \alpha y(t - \tau_a(t);{\theta_0,\theta_c}) + n(t)$, where
\begin{equation}
    \tau_a(t) = \frac{2R_a + 2R(t)}{c}.
\end{equation}
Since the authorized radar receiver knows the antenna selection pattern, it can effectively compensate for the introduced phase noise by keeping track of the selected antennas at each time instance and applying the inverse of the known phase shifts $\phi_p(t,{\theta_0,\theta_c})$ to the mixed signals. Then the receiver can perform a DFT to retrieve the true vital sign frequencies. As a result, the proposed antenna selection schemes can be incorporated directly into the existing phased array radar hardware to provide a layer of privacy protection while no affecting the vital sign monitoring performance. Also, we can see that the designing of $\phi_p(t,{\theta_0,\theta_c})$ is independent with the direction of the eavesdropper. Even though CW is considered throughout the paper, the above schemes are also applicable to the linear FMCW case with slight adjustments.

\section{Numerical Results}

\begin{figure}[t]
\centerline{\includegraphics[width=2.8in]{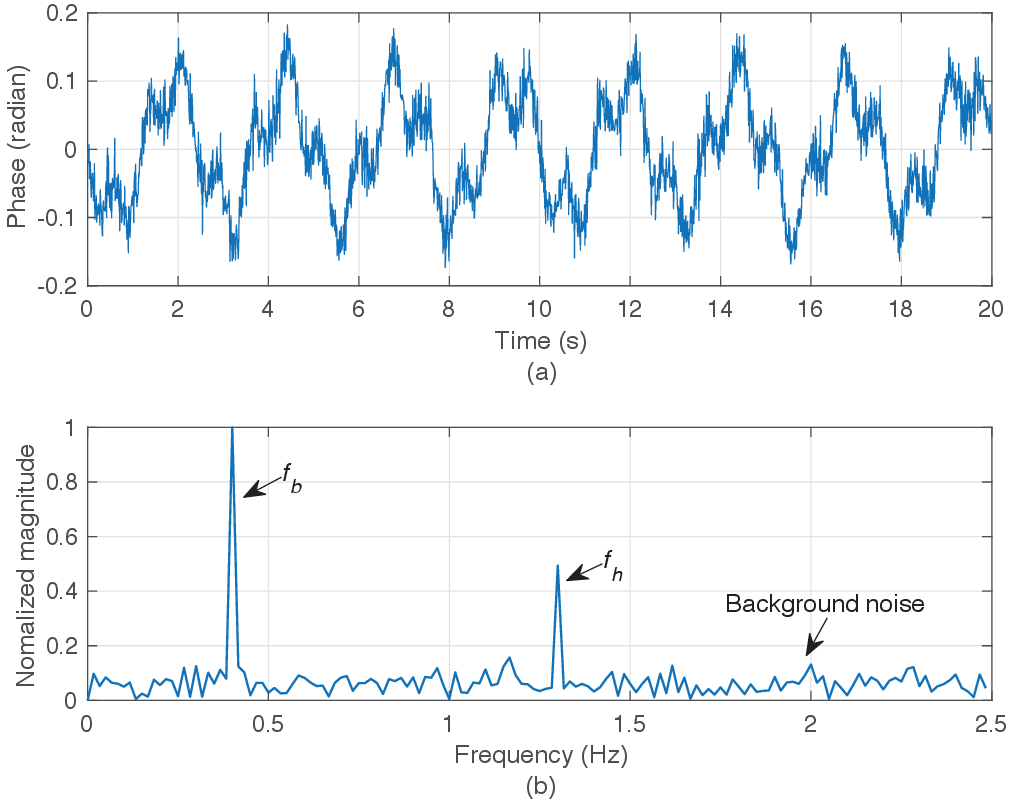}}
\caption{VSM via a conventional phased array: (a) the extracted phases and (b) estimated frequency spectrum of extracted phases by the eavesdropper.}\label{fig3}
\end{figure}

In this section, the privacy protection performance of our proposed antenna selection-enabled radar transmitter is validated via numerical experiments and compared with conventional phased array transmitter. In the following experiments, we set $f_c = 2.2$ GHz, $R_a = R_e = 1$ m, $\theta_0 = 30^{\circ}$, $N = 16$, and the actual $A_h$, $f_h$ as $0.5$ mm, $1.3$ Hz and $A_b$, $f_b$ as $1$ mm, $0.4$ Hz, respectively. The radar transmits a CW signal. The sampling rate of the received signal is set as $100$ Hz, which is sufficient to estimate the low-frequency vital signs ($0.2-2$ Hz). The signal-to-noise ratio (SNR) is set to $10$  dB unless otherwise specified. Moreover, a simulated annealing algorithm is adopted to optimize \eqref{eq5}.

\subsection{Privacy of the Conventional Phased Array Radar Transmitter}
In the first experiment, we adopt a conventional phased array for VSM, to demonstrate its privacy vulnerability. We use the attack method of Section \ref{sec2} to extract the phase of radar echoes and estimate its frequency spectrum. In Fig. \ref{fig3}, the phases of the reflected signals with receiver noise show a nearly periodic pattern over time. This pattern corresponds to chest movements caused by the user’s vital sign activities.
As shown in Fig. \ref{fig3}(b), the two frequency peaks precisely correspond to the true breathing and heartbeat frequencies (0.4 Hz and 1.3 Hz, respectively), indicating that the user's vital sign information is exposed to the eavesdropper and that the conventional phased array lacks privacy protection.

\begin{figure}[t]
\centerline{\includegraphics[width=2.8in]{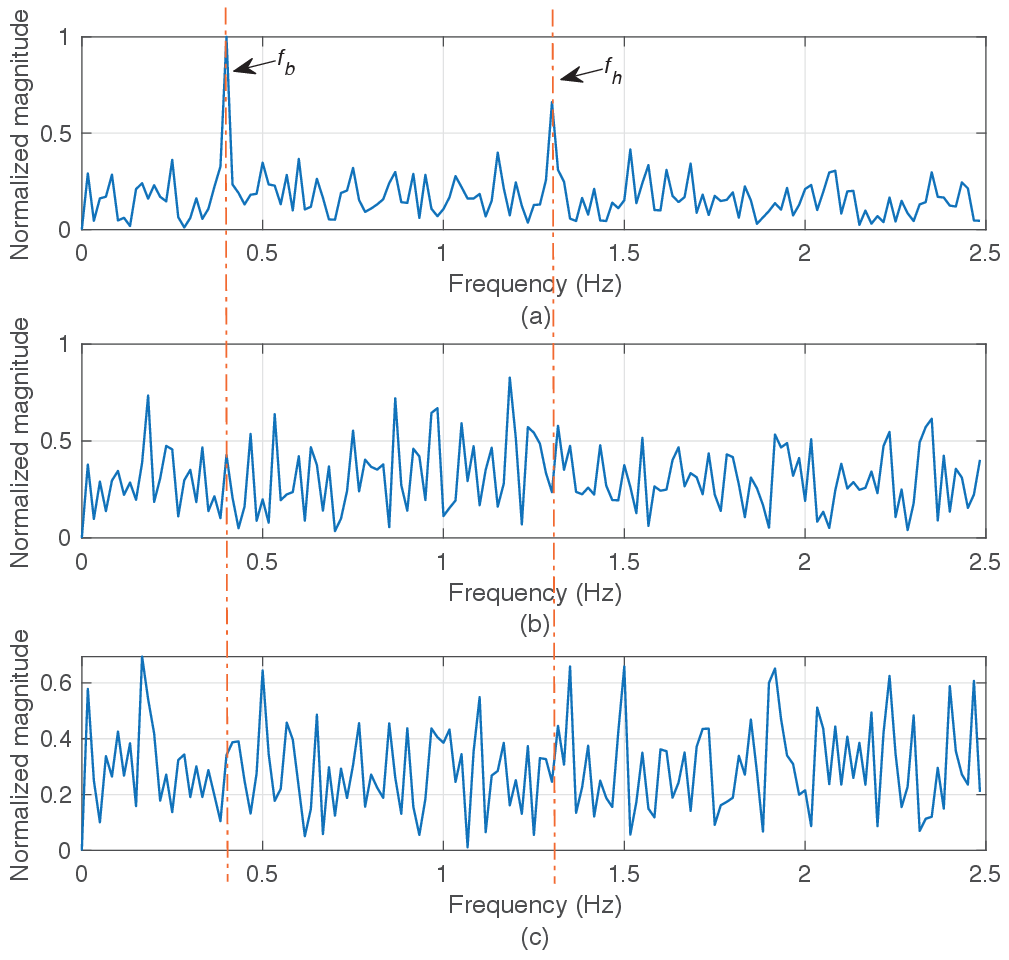}}
\caption{VSM via an antenna selection-enabled phased array: (a) the estimated frequency spectrum of the extracted phases by the authorized receiver and  the estimated frequency spectrum of the extracted phases by the eavesdropper after using (b) MPV-I and (c) MPV-II, respectively.}\label{fig4}
\end{figure}

\subsection{Privacy of the Antenna Selection-Enabled Radar Transmitter}
In the second experiment, the vital sign monitoring and privacy preserving performance of our proposed antenna selection-enabled transmitter are evaluated. For simplicity, the proposed first and second maximum phase variance (MPV) scheme shown in Section \ref{sec3b} are denoted as MPV-I and MPV-II, respectively. In MPV-II, $L = 12$ and $\theta_c = 41^{\circ}$. The performance of vital sign monitoring at the authorized receiver after using antenna selection is shown in Fig. \ref{fig4} (a), where we employ MPV-I to optimize the antenna selection pattern. From Fig. \ref{fig4} (a), we can see that the true vital sign frequencies are extracted accurately as the two highest frequency peaks exactly correspond to the actual $f_b$ and $f_h$. This indicates that the proposed method maintains the integrity of vital sign monitoring for the authorized receiver. A similar monitoring result can be obtained by using MPV-II. The results of privacy enhancing after using MPV-I and MPV-II to optimize the antenna selection pattern are exhibited in Fig. \ref{fig4} (b) and (c), respectively. In both cases, the resulting frequency spectra of the extracted phases by the eavesdropper receiver are highly obfuscated, with no clear peaks corresponding to the true vital sign frequencies. This demonstrates that the phase noise introduced by each antenna selection scheme successfully masks the actual physiological information, preventing an eavesdropper from accurately deciphering the vital sign frequencies.

\begin{figure}[t]
\centerline{\includegraphics[width=2.8in]{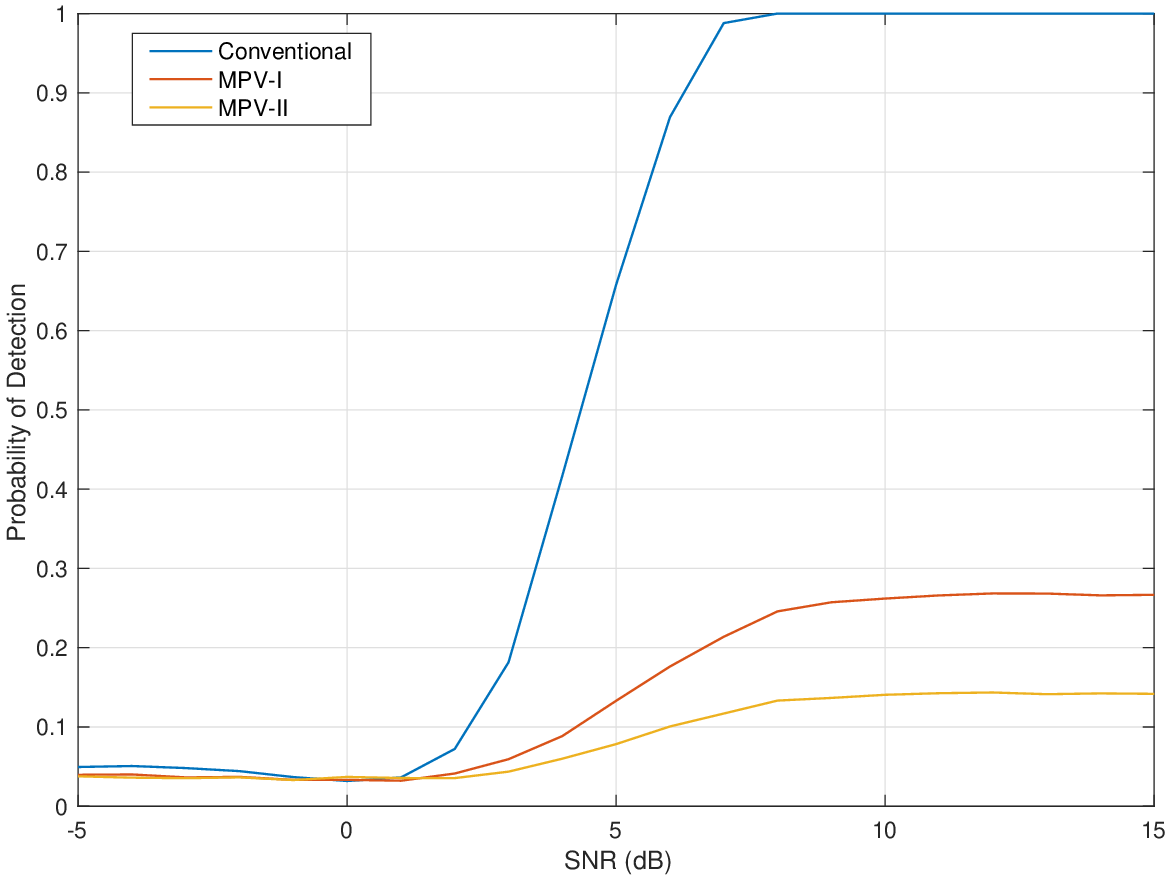}}
\caption{Probability of Detection vs. SNR for the conventional phased array and the proposed antenna selection-enabled ones.}\label{fig5}
\end{figure}

{Next, we evaluate the probability of detection (POD) by the eavesdropper when  the transmitter is a conventional phased array using the proposed antenna selection-enabled schemes, MPV-I and MPV-II. 
{We conducted $N_c=1e4$ Monte Carlo runs; in each run, the eavesdropper extracts the maximum and the second maximal frequency peaks, and checks whether  they are equal to  $f_h$ and $f_b$ in any order.} If they are, a successful detection is declared, and the POD is computed as the number of successful detection over $N_c$.
} The results are shown in Fig. \ref{fig5}, from which we can see that, at low SNRs, the POD for all three setups (conventional, MPV-I, and MPV-II) remains low. This is expected because, in low SNR conditions, the receiver noise  significantly masks the vital sign signals, making it difficult for the eavesdropper to accurately estimate the vital sign frequencies. As the SNR increases, the POD for all methods increases, as the impact of receiver noise diminishes and the vital sign signals become more distinguishable. When the SNR is high, the POD for the conventional phased array converges to 1, indicating that the eavesdropper can reliably detect the true vital sign frequencies with high accuracy, and thus, the conventional phased array fails to provide any privacy protection. In contrast, the PODs for MPV-I and MPV-II remain less than 1 even at high SNR levels, indicating that the artificial phase noise introduced by these antenna selection schemes continues to obscure the true vital sign frequencies and thus reduces the likelihood of successful detection by the eavesdropper. These results further validate that the proposed MPV-I and MPV-II schemes provide effective privacy protection across varying noise levels, whereas the conventional phased array does not.

\section{Conclusion}
We have introduced a privacy-enhancing antenna selection technique for radar-based VSM systems using phased arrays. We have first showcased that the conventional phased array systems are vulnerable to eavesdropping.
Then we have proposed an antenna selection-enabled transmitter that dynamically activates a subset of antennas, introducing time-varying phase shifts. We have also designed two antenna selection patterns to maximize the variance of the introduced phase shifts so as to improve privacy.
Numerical results showed that both approaches effectively mask vital sign frequencies and enhance privacy, while an authorized receiver, with knowledge of the antenna selection pattern, can readily recover the vital signs by compensating for the introduced phase noise. In the future, we will study further the effects of antenna selection on the radar beam pattern and receiver SNR as the designed antenna selection mechanisms affect not only the phases of received radar signal but also its magnitude.

\begin{appendices}
\section{Proof of Proposition 1}
We set up the Lagrangian function of \eqref{eq6} with KKT multipliers as follows
\begin{equation}
    \mathcal{L} = \sum_{i=1}^M p_i \phi_i^2 - \left( \sum_{i=1}^M p_i \phi_i \right)^2 + \gamma \left( 1 - \sum_{i=1}^M p_i \right) + \sum_{i=1}^M \eta _i p_i,
\end{equation}
where the Lagrange multiplier \( \gamma \) is for the equality constraint and the KKT multipliers \( \eta_i \) are for the inequality constraints. We then compute the partial derivative of \( \mathcal{L} \) with respect to \( p_i \):
\begin{equation}
\frac{\partial \mathcal{L}}{\partial p_i} = \phi_i^2 - 2 \phi_i \sum_{j=1}^M p_j \phi_j - \gamma + \eta_i = 0.
\end{equation}
Then, the above equation simplifies to:
\begin{equation}
\phi_i^2 - 2 \phi_i \mu - \gamma = - \eta_i.
\end{equation}
The complementary slackness condition implies that $\eta_i = 0$ when $p_i > 0$ and $\eta_i \geq 0$ when $p_i = 0$, which leads to
\begin{equation}\label{eq7}
\phi_i^2 - 2 \phi_i \mu = \gamma \quad \text{for} \quad p_i>0.
\end{equation}
\eqref{eq7} implies that for all \( i \) with \( p_i > 0 \), \( \phi_i^2 - 2 \phi_i \mu \) must be constant. Next, we can prove easily that there are only two phases with probability larger than 0 by finding that having more than two distinct $\phi_i$ with positive probabilities is contradictory to \eqref{eq7}. Denote these two phases as $\phi _a$, $\phi _b$ and their probabilities as $p_a$ and $1-p_a$, respectively, one can obtain the variance as
\begin{equation}
    \begin{split}
        \text{Var}(\phi_p(t)) &= p_{a} \phi _a^2 + (1 - p_{a}) \phi _b^2 - \left( p_{a} \phi _a + (1 - p_{a}) \phi _b \right)^2\\
        &=p_{a} (1 - p_{a}) (\phi _b - \phi _a)^2.
    \end{split}
\end{equation}
Obviously, we can choose $p_a = \frac{1}{2}$, $\phi_b = \phi _{\text{max}}$ and $\phi_a = \phi_{\text{min}}$ to maximize the variance, where $\phi _{\text{max}}$ and $\phi _{\text{min}}$ are respectively the maximal and minimal phase among $\{\phi_i\}_{i=1,\cdots,M}$ introduced by antenna selection.
\end{appendices}

\bibliography{ref}
\bibliographystyle{IEEEtran}

\end{document}